\documentclass[aps,prd,superscriptaddress,tightenlines,nofootinbib,notitlepage]{revtex4-1}
\usepackage[T1]{fontenc}
\usepackage{microtype}
\usepackage[textsize=scriptsize,backgroundcolor=red!70,linecolor=red]{todonotes}
\usepackage{booktabs}
\usepackage{dcolumn}
\usepackage{amsmath}
\usepackage{amsfonts}
\usepackage{amssymb}
\usepackage{graphicx}
\usepackage{ulem}
\usepackage{hyperref}
\usepackage[all]{hypcap}
\usepackage{paralist}
\usepackage{multirow}
\newcommand{\nue}{\ensuremath{\nu_e}}
\newcommand{\numu}{\ensuremath{\nu_\mu}}

\newcommand{\anumu}{\ensuremath{\bar{\nu}_\mu}}

\usepackage{graphicx}
\usepackage{dcolumn}
\usepackage{bm}
\usepackage{epsf}
\usepackage{dcolumn}
\def\be{\begin{equation}}
\def\ee{\end{equation}}
\def\bea{\begin{eqnarray}}
\def\eea{\end{eqnarray}}

\def\beq{\begin{equation}}
\def\eeq{\end{equation}}

\def\beqa{\begin{eqnarray}}
\def\eeqa{\end{eqnarray}}

\begin{document}

\DeclareGraphicsExtensions{.eps,.ps}

\title{Testing for coherence and nonstandard neutrino interactions in COHERENT data}

\author{Jiajun Liao}
\email[Email Address: ]{liaojiajun@mail.sysu.edu.cn}
\affiliation{School of Physics, Sun Yat-sen University, Guangzhou, 510275, China}
 
\author{Danny Marfatia}
\email[Email Address: ]{dmarf8@hawaii.edu}
\affiliation{Department of Physics and Astronomy, University of Hawaii at Manoa, Honolulu, HI 96822, USA}

\author{Jiajie Zhang}
\email[Email Address: ]{zhangjj253@mail2.sysu.edu.cn}
\affiliation{School of Physics, Sun Yat-sen University, Guangzhou, 510275, China}

\begin{abstract}

We analyze data from the CsI, liquid Ar and Ge detectors of the COHERENT experiment and confirm within $1.5\sigma$  that the measured elastic neutrino-nucleus scattering cross section is proportional to the square of the number of neutrons in the nucleus, as expected for coherent scattering in the standard model. We also show how various degeneracies involving nonstandard neutrino interaction parameters are broken in a combined analysis of the three datasets.

\end{abstract}
\maketitle

\section{Introduction}
 Neutrinos play an important role in exploring physics beyond the standard model (SM), and various methods have been developed to detect them~\cite{ParticleDataGroup:2022pth}. For neutrinos with energy $E_\nu\lesssim 100$ MeV, the dominant neutrino interaction process is  
coherent elastic neutrino-nucleus scattering (CE$\nu$NS). CE$\nu$NS occurs when the momentum transferred in scattering is smaller than the inverse of the radius of the nucleus, so that the scattering amplitudes of nucleons inside the nucleon can be added coherently~\cite{Freedman:1973yd}. This yields a large enhancement in the neutrino cross section. However, the small deposited nuclear energy ($\sim$~keV)  makes CE$\nu$NS difficult to detect.
The first detection of CE$\nu$NS was made by the COHERENT experiment in 2017 with a sodium-doped cesium iodide (CsI[Na]) scintillation detector using neutrinos produced by the Spallation Neutron Source (SNS) at the Oak Ridge National Laboratory~\cite{COHERENT:2017ipa}. 
COHERENT has also detected CE$\nu$NS in a single-phase liquid argon (LAr) detector~\cite{COHERENT:2020iec}, and has updated CsI data with a larger exposure~\cite{COHERENT:2021xmm}. Recently, the COHERENT Ge-Mini detector made a detection of CE$\nu$NS in germanium (Ge). Thus, COHERENT has detected CE$\nu$NS in elements with widely different atomic numbers.
Planned upgrades include ton-scale NaI[Tl], 750~kg LAr, and cryogenic CsI detectors~\cite{COHERENT:2020ybo}.

The measurement of CE$\nu$NS has important implications for particle physics, nuclear physics and astrophysics; for a recent review see Ref.~\cite{Abdullah:2022zue}. In the SM, the CE$\nu$NS  cross section  with nuclear recoil energy $E_r$ from scattering on a nucleus with mass $m_N$ is
\begin{equation}
	\frac{d\sigma}{dE_r}=\frac{G_F^2m_N}{\pi}
	\left(1-\frac{E_r}{E_\nu}-\frac{E_rm_N}{2E_\nu^2}\right)Q_\text{SM}^2\,,
 \label{eq:cevns-xsec}
\end{equation}
where $Q_\text{SM}=Ng_V^nF_N(q^2)+Zg_V^pF_Z(q^2)$ is the SM weak charge with $N$ ($Z$) the number of neutrons (protons) inside the nucleus, $F_N(q^2)$ ($F_Z(q^2$))  the neutron (proton) form factor, and $q$ the momentum transfer. The weak coupling constants to the neutron and proton are $g_V^n=-1/2$ and $g_V^p=1/2-2 \sin^2\theta_W$, respectively. 

Since the weak mixing angle $\sin^2\theta_W\simeq 0.238$ at low energy~\cite{ParticleDataGroup:2022pth}, $g_V^p \ll g_V^n$ and the cross section of CE$\nu$NS is approximately proportional to $N^2$. 
This $N^2$ dependence is a characteristic feature in the SM, and can be tested by measurements of CE$\nu$NS in elements with different neutron numbers. In addition, 
the strong dependence of the CE$\nu$NS cross section on the composition of the nucleus enables measurements in multiple elements to place robust constraints on
nonstandard interactions (NSI) of neutrinos. We perform a combined analysis of all COHERENT datasets to check the $N^2$ dependence of the CE$\nu$NS cross section in the SM. We also examine degenerancies between the NSI and SM parameters, and between the NSI parameters themselves.

The paper is organized as follows. In Section II, we describe our analysis of the CsI, LAr and Ge datasets. We combine the datasets and perform a test of the $N^2$ dependence in Section III. In Section IV, we investigate the constraints on NSI. We summarize our main results in Section V.

\section{Data analysis}

The COHERENT experiment has deployed various detectors in the {\it Neutrino Alley} at the SNS between 19~m and 28~m from the mercury target. 
 Neutrinos detected at the SNS are generated by bombarding a mercury target with a high-energy proton beam, which produces a large number of pions. 
The $\pi^-$ mesons are captured on target nuclei, while
the $\pi^+$ mesons are quickly stopped inside the target, and decay into $\mu^+$ and a prompt $\nu_\mu$  via $\pi^+\to \mu^++\numu$. After traveling one tenth of a millimeter, the $\mu^+$ also decays at rest and produces delayed $\anumu$ and $\nu_e$ via $\mu^+\to e^++\anumu+\nue$. The distribution of the neutrino flux at SNS is given by
 \begin{equation}
    \begin{aligned}
    &\frac{dN_{\nu_\mu}}{dE_\nu}(E_\nu)=\mathcal{N} \delta\left(E_\nu-\frac{m_\pi^2-m_\mu^2}{2m_\pi}\right),\\
    &\frac{dN_{\bar{\nu}_\mu}}{dE_\nu}(E_\nu)=\mathcal{N}\frac{64E_\nu^2}{m_\mu^3}\left(\frac34-\frac{E_\nu}{m_\mu}\right),\\
    &\frac{dN_{\nu_e}}{dE_\nu}(E_\nu)=\mathcal{N}\frac{192E_\nu^2}{m_\mu^3}\left(\frac12-\frac{E_\nu}{m_\mu}\right)\,,
    \end{aligned}
\label{eq:flux}
\end{equation}
where $\mathcal{N} = \frac{r N_{\text{POT}}}{4\pi L^2}$ is a normalization factor with $L$ the distance between the detector and the SNS target, and $N_{\text{POT}}$ the total number of protons on target (POT), which can be calculated from the beam power and energy. The parameter $r$ represents the number of neutrinos per flavor per POT. At the SNS, the proton beam energy has been slightly increased from 1~GeV to 1.3~GeV~\cite{COHERENT:2020ybo}, and the operating power has been enhanced from 1.4~MW to 1.7~MW~\cite{Adamski:2024yqt}, with prospects for further upgrades to 2.8 MW in the future~\cite{2.8MW}. 
A higher proton beam power produces a higher neutrino yield.

The number of CE$\nu$NS events of each flavor $\ell$ in the $i^\text{th}$ energy bin and  $j^\text{th}$ time bin can be calculated by the convolution of the neutrino flux and the differential cross section,
\begin{equation}
    N_{ij,\ell}=n_{det} \int_{t_{j}}^{t_{j+1}}dt P_{\ell}(t) \int_{E_r^{i}}^{E_r^{i+1}}dE_r\epsilon(E_r)\int_{{E_{\nu}^{min}}}^{E_{\nu}^{max}}dE_{\nu}\frac{dN_{\ell}}{dE_{\nu}}(E_{\nu})\frac{d\sigma}{dE_r}(E_{\nu},E_r)\,,
    \label{eq:events}
\end{equation}
where $\epsilon(E_r)$ is the detector efficiency and $n_{\text{det}} = \frac{m_{\text{det}} N_A}{M}$ with $N_A$ the Avogadro's constant, $m_{\text{det}}$ the mass of the detector, and $M$ the molar mass number of the detector element. In our analysis, we assume that the neutron form factor is the same as the proton form factor, and adopt the Klein-Nystrand form factor~\cite{Klein:1999qj}. Other choices of form factors yield a similar result~\cite{AristizabalSierra:2019zmy}.
To calculate the number of CE$\nu$NS events in each time bin, we use the time distribution probability function $P_{\ell}(t)$ at the SNS for the CsI and LAr detectors~\cite{Konovalov}. Note that for the Ge-mini detector, we incorporate the signal time distribution due to the drift times within the large diodes~\cite{citation-key2024}. We detail the analysis for each detector below.

\subsection{CsI}
The COHERENT CsI detector is composed of a 14.6~kg CsI[Na] crystal, and is located at a distance of 19.3~m from the mercury target. The updated dataset has a total integrated beam power of 13.99~GWhr, which corresponds to $N_{\text{POT}} = 3.2 \times 10^{23}$~\cite{COHERENT:2021xmm}. The proton beam was operated at a mean energy of 0.984 GeV, which yields $r = 0.0848$. To convert the observed number of photoelectrons (PE) to the nuclear recoil energy, we adopt the detection efficiency and quenching factor given in Ref.~\cite{COHERENT:2021xmm}.
The energy distribution is divided into 8~bins ranging from 8~PE to 60~PE, and the time distribution is divided into 11~bins spanning 0 to 6 $\mu$s. To evaluate the statistical significance, we define
\begin{equation}
\label{chi2-csi}
\chi_{\text{CsI}}^{2}=2\sum_{i=1}^8{\sum_{j=1}^{11}{}}\left[ N_{\text{th},ij}^{\text{CsI}}-N_{ij}^{\exp}+N_{ij}^{\exp}\ln \left( \frac{N_{ij}^{\exp}}{N_{\text{th},ij}^{\text{CsI}}} \right) \right] +\sum_{k=0}^{3}\left( \frac{\alpha _k}{\sigma _k} \right) ^2\,,
\end{equation}
where
$N_{\text{th},ij}^{\text{CsI}} = \left(1 + \alpha_0\right) N_{ij}^{\text{CsI}} + \left(1 + \alpha_1\right) N_{ij}^{\text{BRN}} + \left(1 + \alpha_2\right) N_{ij}^{\text{SSB}} + \left(1 + \alpha_3\right) N_{ij}^{\text{NIN}}$ 
with $N_{ij}^{\text{CsI}}$ the predicted total number of CE$\nu$NS events from Cs and I. 
We take the experimentally observed counts $N_{ij}^{\text{exp}}$, the Steady State Background $N_{ij}^{\text{SSB}}$, the Beam-Related Neutrons $N_{ij}^{\text{BRN}}$, and Neutrino-Induced Neutrons $N_{ij}^{\text{NIN}}$ from the data release of Ref.~\cite{COHERENT:2021xmm}.
Here, $\sigma_0 = 11.45\%$ is the normalization uncertainty that includes the efficiencies, neutrino flux and quenching factor uncertainties, $\sigma_1 = 25\%$ is the uncertainty from BRN, $\sigma_2 = 2.1\%$ is the uncertainty from the SSB, and $\sigma_3 = 35\%$ corresponds to the uncertainty associated with NIN~\cite{COHERENT:2021xmm}. The parameters $\alpha_0$, $\alpha_1$, $\alpha_2$, and $\alpha_3$ are nuisance parameters for CE$\nu$NS, BRN, SSB, and NIN events, respectively. 

\begin{figure}
	\centering
	\includegraphics[width=0.45\textwidth]{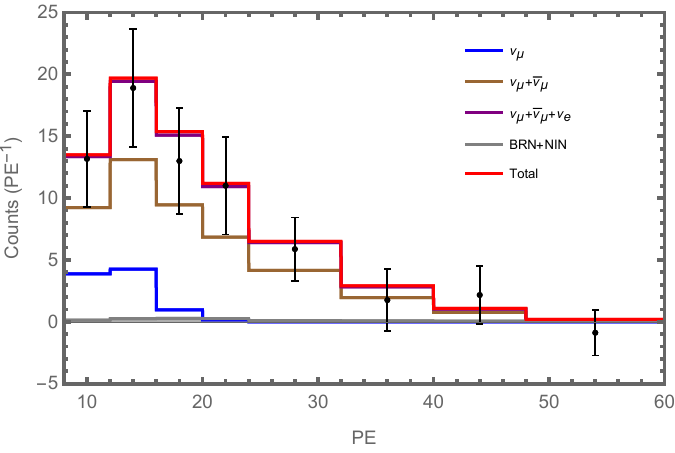}
	\includegraphics[width=0.45\textwidth]{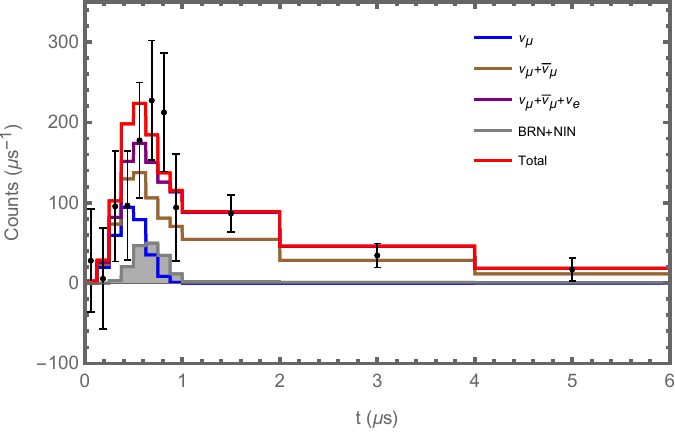}
	\caption{SM expectations for the energy (left) and time (right) distributions are in good agreement with COHERENT-CsI data.} 
	\label{fig:csi_events}
\end{figure}

We find the SM best fit gives $\chi^2_\text{CsI, min}=95.9$ with $\alpha_{0} = -0.024$, $\alpha_{1} = 0.047$, $\alpha_{2} = -0.018$, and $\alpha_{3} = -0.012$. We show the energy and time distributions in Fig.~\ref{fig:csi_events}, respectively. It is evident that the SM is in good agreement with COHERENT-CsI data.

\subsection{LAr}
The CENNS-10 detector is composed of 24~kg of LAr and is located 27.5~meters from the center of the mercury target. The latest release of data collected between July 2017 and December 2018 with a total integrated beam power of 6.12~GWhr, corresponds to $N_{\text{POT}} = 1.37 \times 10^{23} $~\cite{COHERENT:2020iec}. During this period the beam energy was 1~GeV and the beam power was 1.4~MW, which yields the same $r$ as the CsI detector~\cite{COHERENT:2020iec}.  The detection efficiency and quenching factor for the CENNS-10 detector are given in Ref.~\cite{COHERENT:2020iec}.
The energy and time distributions of the CENNS-10 data are divided into 12 and 10 bins, respectively, which give a total of 120 bins.  We define
\begin{equation}
\label{chi2-lar}
	\chi _{\text{LAr}}^{2}=\sum_{i=1}^{12}{\sum_{j=1}^{10}{\left( \frac{N_{ij}^{\exp}-\sum_{k=0}^3{\left( 1+\beta _k+\sum_n{\eta _{k;n} \Delta N_{kn;ij}^{\rm {sys}}} \right)}N_{ij}^{k}}{\sigma _{ij}} \right)}}^2+\sum_{k=0}^3{\left( \frac{\beta _k}{\sigma _k} \right)}^2+\sum_{k,n}\eta_{kn}^2\,,
\end{equation}
where $k=0,1,2,3$ corresponds to the CE$\nu$NS, prompt beam-related neutrons (pBRN), delayed beam-related neutrons (dBRN), and SSB, respectively. 
The uncertainty in each bin is $\sigma_{ij}=\sqrt{N_{ij}^\text{exp}+N_{ij}^\text{SSB}/5}$, and the uncertainties for $\sigma_0$(CE$\nu$NS),  $\sigma_1$(pBRN),  $\sigma_2$(dBRN), and  $\sigma_3$(SSB) are taken to be 0.13, 0.32, 1.0, and 0.0079, respectively~\cite{COHERENT:2020iec}. $\beta_k$ are nuisance parameters.
We account for shape uncertainties in the CE$\nu$NS signal and the pBRN background by introducing $\Delta N_{kn;ij}^{\rm {sys}} \equiv \frac{N_{kn;ij}^{\rm{sys}} - N_{kn;ij}^\text{CV}}{N_{kn;ij}^\text{CV}}$, where $n$ is the index of uncertainty associated with the $k^\text{th}$ component in each bin, and $N_{kn;ij}^\text{CV}$ denotes the central value (CV) of the CE$\nu$NS or pBRN distribution~\cite{AtzoriCorona:2022qrf, DeRomeri:2022twg}. 
The CE$\nu$NS shape uncertainties arise from the mean time to trigger distribution, and the energy distributions of the pulse-shape discrimination parameter $F_{90}$, which is the ratio of the fast signal from the first 90~ns of the pulse to the total number of PE in a 6~$\mu s$ window after the initial pulse. The pBRN shape uncertainties mainly come from the energy distribution, the mean time to trigger distribution, and the trigger width distribution. 
The experimental data $N_{ij}^{k}$, the systematic probability distribution functions $N_{kn;ij}^{\rm {sys}}$, the central values $N_{kn;ij}^{\rm {CV}}$ and the distributions related to the shape uncertainties are taken from the data release of Ref.~\cite{COHERENT:2020ybo}.

\begin{figure}
	\centering
	\includegraphics[width=0.45\textwidth]{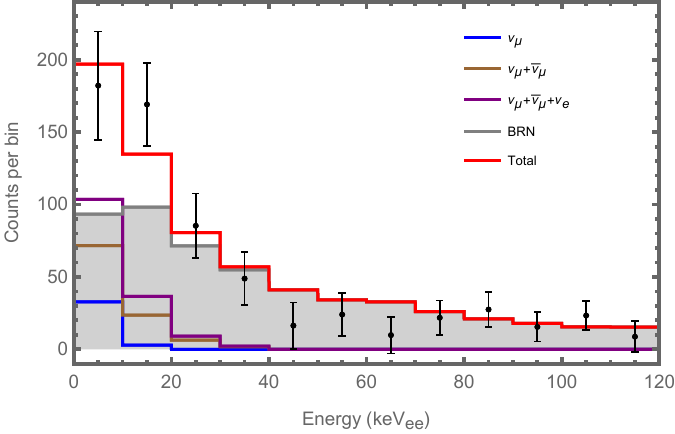}
	\includegraphics[width=0.45\textwidth]{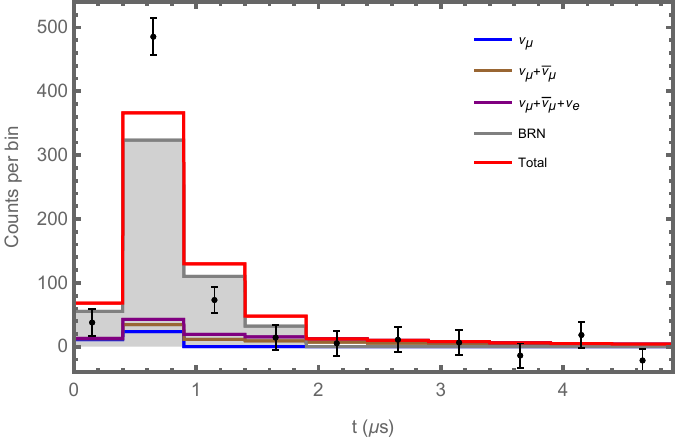}
	\caption{Same as Fig.~\ref{fig:csi_events} except for COHERENT-LAr. }
	\label{fig:lar_events}
\end{figure}

For the SM best fit we find $\chi^2_\text{LAr, min}=107.2$ with $\beta_{0} = 0.0004$,  $\beta_{1} = 0.1538$, $\beta_{2} = -0.578$ and $\beta_{3} = -0.0035$. The energy and time distributions are shown in Fig.~\ref{fig:lar_events}. The SM expectation is easily compatible with the COHERENT-LAr dataset which has lower statistics than the CsI dataset. 

\begin{figure}
	\centering
	\includegraphics[width=0.45\textwidth]{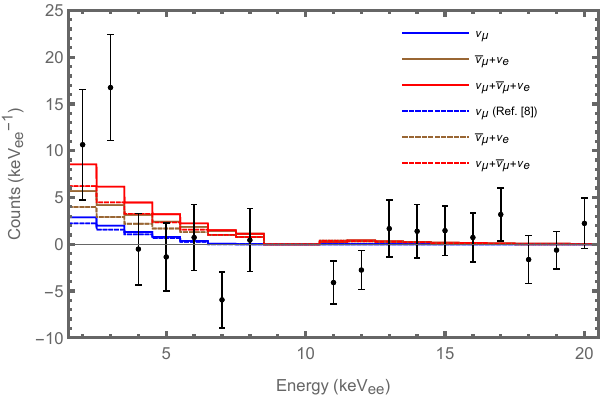}
	\includegraphics[width=0.45\textwidth]{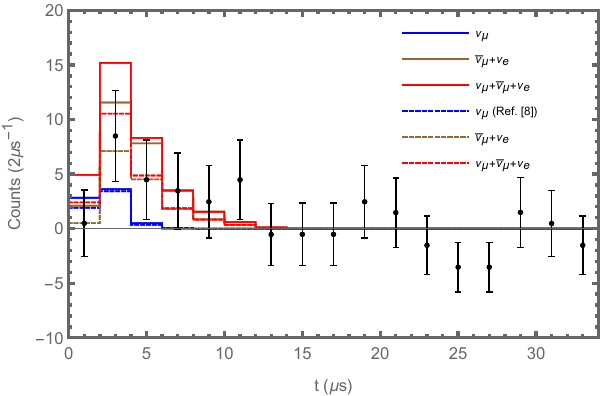}
	\caption{The energy and time distributions of the COHERENT-Ge data. The solid histograms represent our best-fit result with $\gamma_{0} = -0.165$, while the dashed histograms are from Ref.~\cite{Adamski:2024yqt}. The observed number of events and their uncertainties are extracted from the background-subtracted spectra in time and energy from the on-beam data in Ref.~\cite{Adamski:2024yqt}. 
    }
	\label{fig:Ge_events}
\end{figure}

\subsection{Ge}
The Ge-Mini detector located 19.2~m from the SNS target is composed of eight low-background cryostats, each housing an Inverted Coaxial Point-Contact (ICPC) detector with a weight of 2.2~kg.  After excluding the inactive surface layer of the diodes, the effective mass of the Ge-Mini detector is 10.66 kg~\cite{Adamski:2024yqt}. With the beam power increased to 1.7~MW, the total exposure is 10.22~GWhkg. The number of neutrinos per flavor per POT, corresponding to a proton beam energy of 1.05~GeV, is 0.096~\cite{Adamski:2024yqt}. 

Since only the electronic ionization is measured at the Ge spectrometers, a quenching factor has to be introduced to convert the nuclear recoil energy to the observed ionization energy.
We employ the Lindhard model~\cite{Lindhard} to describe the quenching factor:
\begin{equation}
Q(E_r)=\frac{kg(\epsilon)}{1+kg(\epsilon)}\,,
\end{equation}
where $k=0.133 Z^{2/3} A^{-1/2}\approx0.157$ measures the electronic energy loss in natural Ge, and $g(\epsilon)=3\epsilon^{0.15}+0.7\epsilon^{0.6}+\epsilon $ with $\epsilon=11.5Z^{-\frac{7}{3}}E_{r}$. Here, $E_r$ is in keV so that $\epsilon$ is dimensionless. The data have a threshold of 1.5~keV$_{\text{ee}}$, which corresponds to a nuclear recoil energy of 6.7~keV$_{\text{nr}}$ with the Lindhard quenching factor. Natural germanium is composed of five stable isotopes, $^{70}$Ge, $^{72}$Ge, $^{73}$Ge, $^{74}$Ge and $^{76}$Ge, with relative abundances, 20.5\%, 27.4\%, 7.76\%, 36.5\% and 7.75\%, respectively. We use the average atomic mass of 72.7 for simplicity. We find that the total number of CE$\nu$NS events is 35.2, which agrees with the SM expectation in Ref.~\cite{Adamski:2024yqt}, and deviates by approximately 2$\sigma$ from the experimental observation of $20.6^{+7.1}_{-6.3}$~\cite{Adamski:2024yqt}.

To calculate the statistical significance of the Ge-Mini data, we define 
\begin{equation}
\label{chi2-ge}
\chi_{\text{Ge}}^{2}=\sum_{i=1}^{17}\left[\frac{N_{i}^\text{exp}-(1 + \gamma_{0})N_{i}^\text{Ge}}{\sigma_i}\right]^{2}+\sum_{j=1}^{17}\left[\frac{N_{j}^\text{exp}-(1 + \gamma_{0})N_{j}^\text{Ge}}{\sigma_j}\right]^{2}+\left(\frac{\gamma_{0}}{\sigma_{0}}\right)^{2}\,,
\end{equation}
where $N_i^\text{Ge}$ ($N_j^\text{Ge}$) is the expected number of CE$\nu$NS events in each energy (time) bin,  
$N_{i}^\text{exp}$ ($N_{j}^\text{exp}$) is the measured number of events in the energy (time) bin, and $\sigma_i$ ($\sigma_j$) is the corresponding uncertainty. Note that energy range [8.5, 11]~keV$_{\text{ee}}$ is excluded from the analysis to mitigate the impact of cosmogenic-induced background lines~\cite{Adamski:2024yqt}, and we only consider data points with $t>0$.
The associated uncertainty $\sigma_{0} = 10.3\%$ incorporates contributions from uncertainties in the flux, detector distance, energy calibration, active mass, and form factor \cite{Adamski:2024yqt}. 

The SM best fit occurs for $\gamma_0 = -0.165$ and has
$\chi^2_\text{Ge, min}=38.4$. The best-fit CE$\nu$NS spectra are shown in Fig.~\ref{fig:Ge_events}, respectively. For comparison, we also show the results obtained from a likelihood fit that utilizes the total signal and total background counts as free parameters~\cite{Adamski:2024yqt}.
Since our best fit is based on the number of events expected in the SM, the counts are higher than in Ref.~\cite{Adamski:2024yqt}. 


\begin{figure}
\centering
\includegraphics[width=0.45\textwidth]{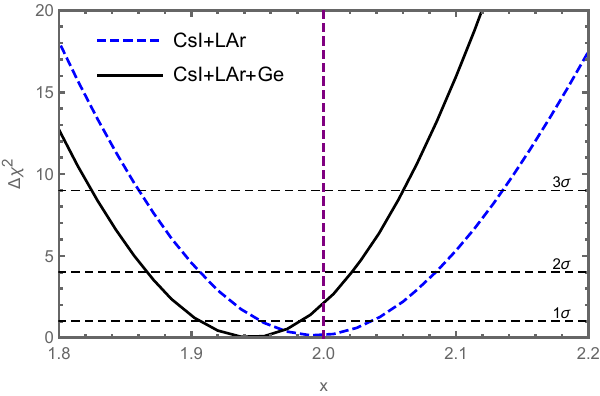}
\caption{$\Delta \chi^2$ as a function of the exponent in Eq.~(\ref{exponent}). The dashed line corresponds to the combined analysis from the CsI and LAr datasets and the solid line corresponds to the combined analysis of the CsI, LAr and Ge datasets. }
\label{fig:N2}
\end{figure}

\section{A test of the $N^2$ dependence}
A salient feature of the SM CE$\nu$NS cross section is its proportionality to $N^2$. This can be understood from the proportionality of the cross section in Eq.~(\ref{eq:cevns-xsec}) to the weak charge squared,
\begin{align}
		Q_\text{SM}^2 &=\left[N{g_V^n}{F_N(q^2)}+Z{g_V^p}{F_Z(q^2)}\right]^2 \\\nonumber
		&= N^2{(g_V^n)}^2{F_N^2(q^2)} + 2NZ{g_V^n}{g_V^p}{F_N(q^2)}{F_Z(q^2)} + Z^2{(g_V^p)}^2{F_Z^2(q^2)}\,.
\end{align}
Since $g_V^n{\gg}g_V^p$ in the SM, the second term is smaller than $8\%$ of the first term for the elements of interest, and the third term is negligible. Therefore, the CE$\nu$NS cross section is approximately proportional to $N^2$. 
Now that the COHERENT experiment has measured CE$\nu$NS in four elements, Cs, I, Ar, and Ge, with sufficiently different neutron numbers, it is possible to test the \( N^2 \) dependence of CE$\nu$NS cross section in the SM. To carry out this test we allow the exponent of the SM weak charge to be a free parameter $x$, i.e., we replace $Q_\text{SM}^2$ in Eq.~(\ref{eq:cevns-xsec}) by 
\begin{align}
	Q^x_\text{SM} =\left[N{g_V^n}{F_N(q^2)}+Z{g_V^p}{F_Z(q^2)}\right]^x\,,
 \label{exponent}
\end{align}
and check how far the combination of the datasets allows $x$ to deviate from 2. Our procedure implicitly assumes that $|x-2|$ can not be too large, in which case the cross terms in the expansion of $Q^x_\text{SM}$ would be sizable, and so we would not be testing the \( N^2 \) dependence.  In Fig.~\ref{fig:N2}, we show $\Delta \chi^2= \chi^2-\chi^2_{\rm min}$ as a function of $x$ for the combined analysis with and without COHERENT-Ge data. The best fit with (without) Ge data occurs for $x=1.95$ ($x=1.99$). We select the Ge dataset for special treatment because, of the three datasets, it is least compatible with the SM. From Fig.~\ref{fig:N2}, it can be seen that the CsI+LAr data are almost perfectly consistent with the $N^2$ dependence, while the combined data from the three detectors are consistent with $x=2$ within 1.5$\sigma$.
The reason that the combined data prefer a smaller value of $x$ is that COHERENT-Ge measured a small number of events than the SM expectation.

\begin{figure}
\centering
\includegraphics[width=0.45\textwidth]{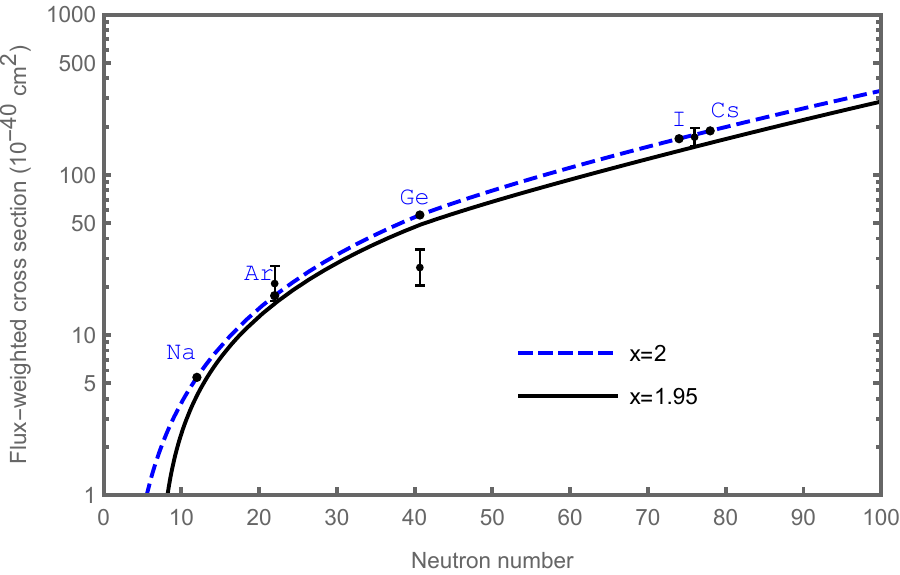}
\caption{The flux-weighted CE$\nu$NS cross section as a function of neutron number. }
\label{fig:cross section}
\end{figure}

To better illustrate the relationship between the CE$\nu$NS cross section and the number of neutrons, we plot the flux-weighted CE$\nu$NS cross section,
\begin{equation}
    \langle \sigma \rangle_\Phi= \frac{1}{3}\sum_{\ell}\int_{0}^{E_r^{\rm max}}dE_r\int_{0}^{E_{\nu}^{\rm max}}dE_{\nu}\frac{1}{\mathcal{N}}\frac{dN_{\ell}}{dE_{\nu}}(E_{\nu})\frac{d\sigma}{dE_r}(E_{\nu},E_r)\,,
    \label{eq:flux-weighted cross section}
\end{equation}
as a function of $N$ in Fig.~\ref{fig:cross section}. 
Here, $\frac{dN_{\ell}}{dE_{\nu}}$ is given by Eq.~(\ref{eq:flux}).
The flux-weighted SM cross sections for Na, Ar, Ge, I, and Cs, are plotted as black points in Fig.~\ref{fig:cross section}. We interpolate these points to obtain the blue dashed curve for $x=2$ in Fig.~\ref{fig:cross section}.
We also show the case with $x = 1.95$ as the black solid curve by applying the same procedure. It should be noted that since the flux-weighted cross section is not only related to the number of neutrons but also to the number of protons and the mass of the nucleons, an exact relationship between the flux-weighted cross section and the number of neutrons cannot be established. The curve obtained by interpolation is only correct at these five points, and there may be deviations at other points.

\begin{table}[t]
\begin{centering}
\begin{tabular}{|c|c|c|c|}
\hline 
& LAr & Ge & CsI  \tabularnewline
\hline 
$\left\langle\sigma\right\rangle_\text{SM}\times10^{-40}\text{cm}^2$ & $17.6$ & $56.1$ & $178$ 
\tabularnewline
\hline 
$\left\langle\sigma\right\rangle_\text{meas}\times10^{-40}\text{cm}^2$ & $20.9^{+6.72}_{-6.86}$ & $26.37^{+8.98}_{-8.58}$ & $172^{+25.8}_{-24.9}$ 
\tabularnewline
\hline 
\end{tabular}
\end{centering}
\caption{The average expected CE$\nu$NS cross section in the SM and the values measured by the LAr, Ge and CsI detectors. }
\label{tab:xsection}
\end{table}

To obtain an average cross section measured by each detector, we multiply the CE$\nu$NS cross section by a constant factor and use the relevant $\chi^2$ function in the previous section to fix it. Then, the average cross section is determined by the flux-weighted CE$\nu$NS cross section multiplied by the measured factor. The average cross sections for Ar, Ge and CsI in the SM and those measured by the three detectors at COHERENT are listed in Table~\ref{tab:xsection}. Except for Ge, the measured average cross sections are consistent with the SM expectations within $1\sigma$.  We also show the average CE$\nu$NS cross sections in Fig.~\ref{fig:cross section}. It is not surprising that the solid curve for $x=1.95$ shows better agreement with the data points than does the dashed curve for $x=2$. 
Note that Fig.~\ref{fig:cross section} is for illustration only since the measured cross section depends on the energy threshold and efficiency of the detector, while the
flux-weighted cross section does not depend on these and is averaged over neutrino energy.
%


\section{Nonstandard neutrino interactions}
We study how the three COHERENT datasets constrain new physics in the NSI framework, which can be expressed as four-fermion contact interactions~\cite{Proceedings:2019qno}:
\begin{equation}
\mathcal{L}_\text{NSI}=-2\sqrt{2}G_F\sum_{q,\ell,\ell^{\prime}}\varepsilon_{\ell\ell^{\prime}}^{qV,C}(\bar{\nu}_\ell\gamma^\mu P_L\nu_{\ell^{\prime}})(\bar{q}\gamma_\mu P_C q)\,,
\end{equation}
where $C=L,R$, and $\varepsilon_{\ell\ell^{\prime}}^{qV}\equiv \varepsilon_{\ell\ell^{\prime}}^{qV, L}+\varepsilon_{\ell\ell^{\prime}}^{qV,R}$ is the neutrino-quark coupling constant in units of $G_F$ with $q=\{u, d\}$ and $\ell,\ell^{\prime}=\{e, \mu,\tau\}$. The weak charge squared in Eq.~(\ref{eq:cevns-xsec}) becomes
\begin{equation}
Q_{\text{NSI},\ell}^2 =\left[\left(g_V^p+2\varepsilon_{\ell\ell}^{uV}+\varepsilon_{\ell\ell}^{dV}\right)Z+\left(g_V^n+\varepsilon_{\ell\ell}^{uV}+2\varepsilon_{\ell\ell}^{dV}\right)N\right]^2+\sum_{\ell,\ell'}\left[\left(2\varepsilon_{\ell\ell'}^{uV}+\varepsilon_{\ell\ell'}^{dV}\right)Z+\left(\varepsilon_{\ell\ell'}^{uV}+2\varepsilon_{\ell\ell'}^{dV}\right)N\right]^2\,.
\label{eq:Q_NSI}
\end{equation}

\subsection{Degeneracies in the weak mixing angle and NSI parameters}

\begin{figure}
	\centering
	\includegraphics[width=0.45\textwidth]{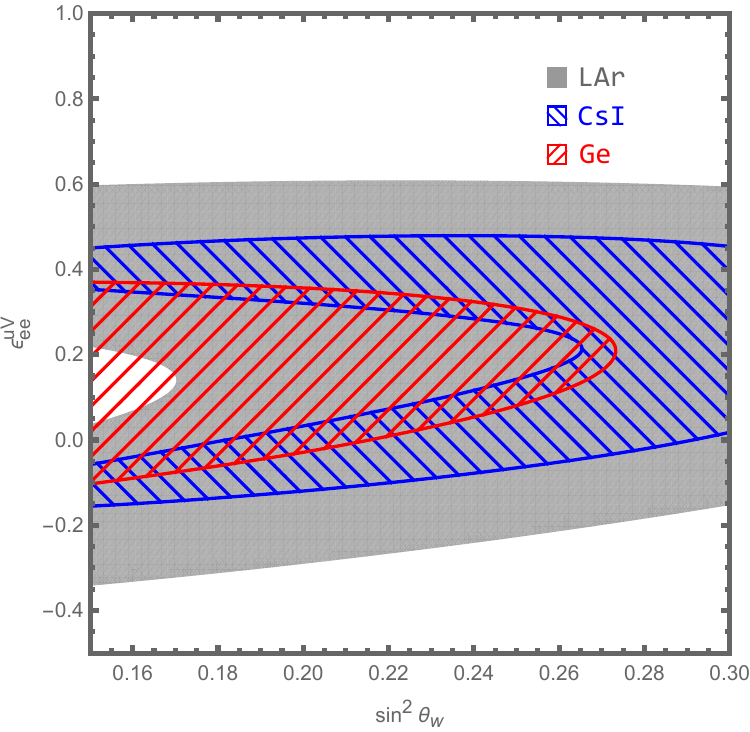}
	\includegraphics[width=0.45\textwidth]{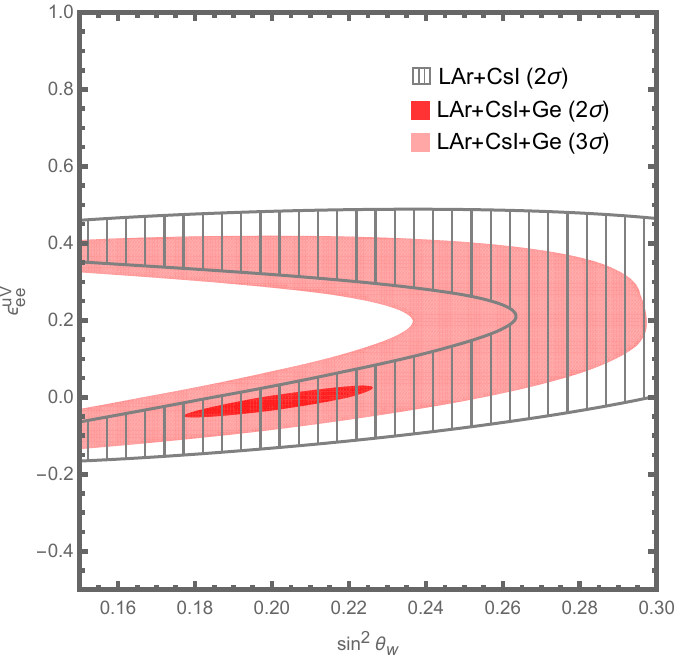}
	\includegraphics[width=0.45\textwidth]{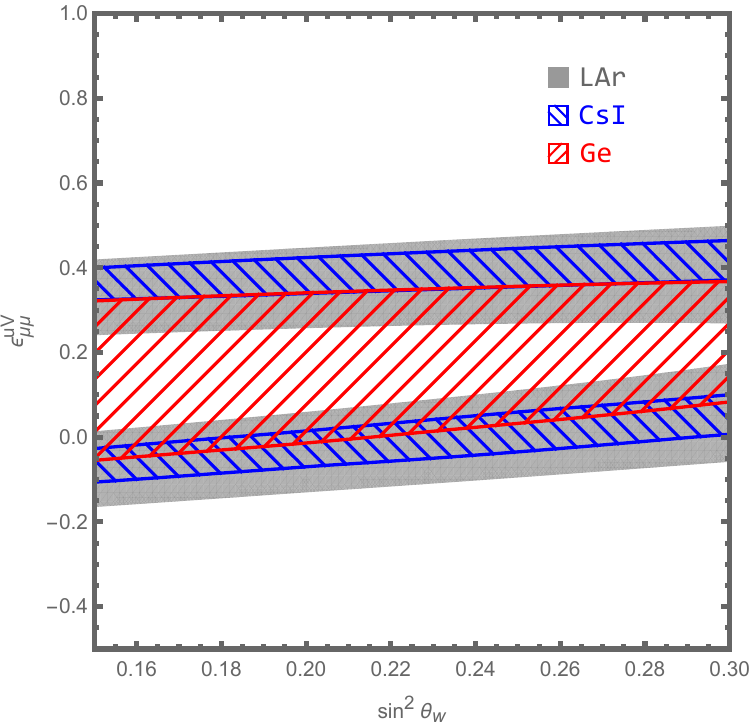}
	\includegraphics[width=0.45\textwidth]{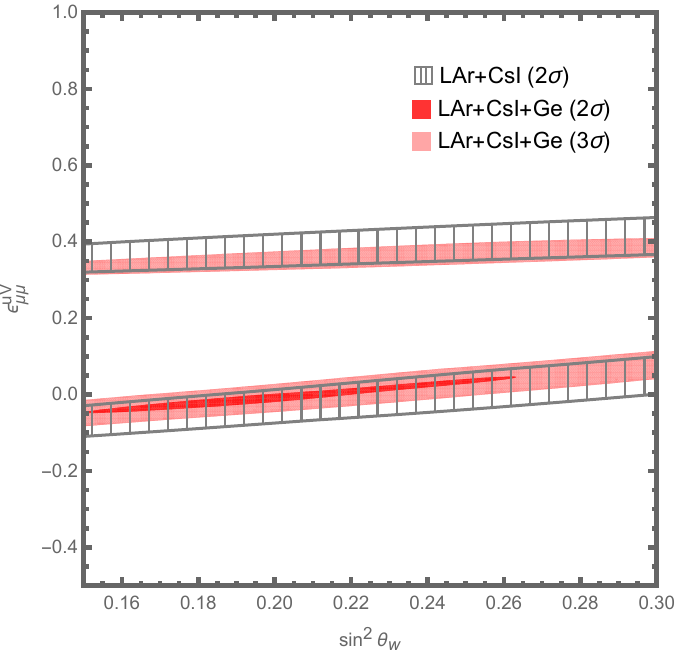}
	\caption{Left panel:  90\%~CL allowed regions in the ($\sin^2\theta_W, \epsilon_{ee}^{uV}$) and ($\sin^2\theta_W, \epsilon_{\mu\mu}^{uV}$) planes from COHERENT LAr, CsI, and Ge data. Right panel: 2$\sigma$ and 3$\sigma$ allowed regions from the combined analysis of the three datasets. The 2$\sigma$ allowed regions from a joint analysis of only CsI and LAr data are shown to emphasize the impact of the Ge data.} 
\label{fig:ee_uu_sinw}
\end{figure}

Several low energy experiments~\cite{Kumar:2013yoa}
have greater sensitivity to $\theta_W$ than CE$\nu$NS. Our purpose is to
show that even in the presence of NSI, the sensitivity of CE$\nu$NS to $\theta_W$ is much improved by combining the data from different elements.
Equation~(\ref{eq:Q_NSI}) shows the existence of degeneracies in $\theta_W$ and the NSI parameters. 
In the case that only $\epsilon_{ee}^{uV}$ or 
$\epsilon_{\mu\mu}^{uV}$ is nonzero, the weak charge can be simplified to
\begin{align}
\label{nsi-sinw2}
Q_{\text{NSI},\ell} &=(0.5-2 \sin^2\theta_W+2\varepsilon_{\ell\ell}^{uV})Z+(-0.5+\varepsilon_{\ell\ell}^{uV})N\,,
\end{align}
where $\ell=e,\mu$.

The parameter regions allowed by the three COHERENT datasets in isolation (left panels) and in combination 
(right panels) in the ($\sin^2\theta_W$, $\epsilon_{ee}^{uV}$) and ($\sin^2\theta_W$, $\epsilon_{\mu\mu}^{uV}$) planes  are shown in the upper and lower panels of Fig.~\ref{fig:ee_uu_sinw}, respectively. 
There are no $1\sigma$ allowed regions in the right panels because the Ge data are slightly discrepant with the SM expectation.
The 3$\sigma$ allowed regions from the combined analysis of the three datasets are comparable in size to the 2$\sigma$ allowed regions without the Ge data.

From Eq.~(\ref{nsi-sinw2}), it is evident that 
$\epsilon_{ee}^{uV}$ ($\epsilon_{\mu\mu}^{uV}$) only affects the detection of $\nue$ ($\numu$/$\anumu$), while all neutrino flavors are affected by $\sin^2\theta_W$. Note that CE$\nu$NS experiments that use nuclear reactors as sources only detect $\bar{\nu}_e$. Then $\epsilon_{ee}^{uV}$ and $\sin^2\theta_W$ are strictly degenerate, and the allowed regions appear as two straight bands~\cite{Galindo-Uribarri:2020huw}.
Since COHERENT detects neutrinos from stopped pions that contain both $\nue$ and $\numu$/$\anumu$, the exact degeneracy between $\epsilon_{ee}^{uV}$ ($\epsilon_{\mu\mu}^{uV}$) and $\sin^2\theta_W$ can be broken.  
The right panels of Fig.~\ref{fig:ee_uu_sinw} show that a combination of the three datasets break the degeneracy between $\epsilon_{ee}^{uV}$ ($\epsilon_{\mu\mu}^{uV}$) and $\sin^2\theta_W$ at 2$\sigma$.

\subsection{Degeneracies in the NSI parameters}
\begin{figure}
\centering
\includegraphics[width=0.45\textwidth]{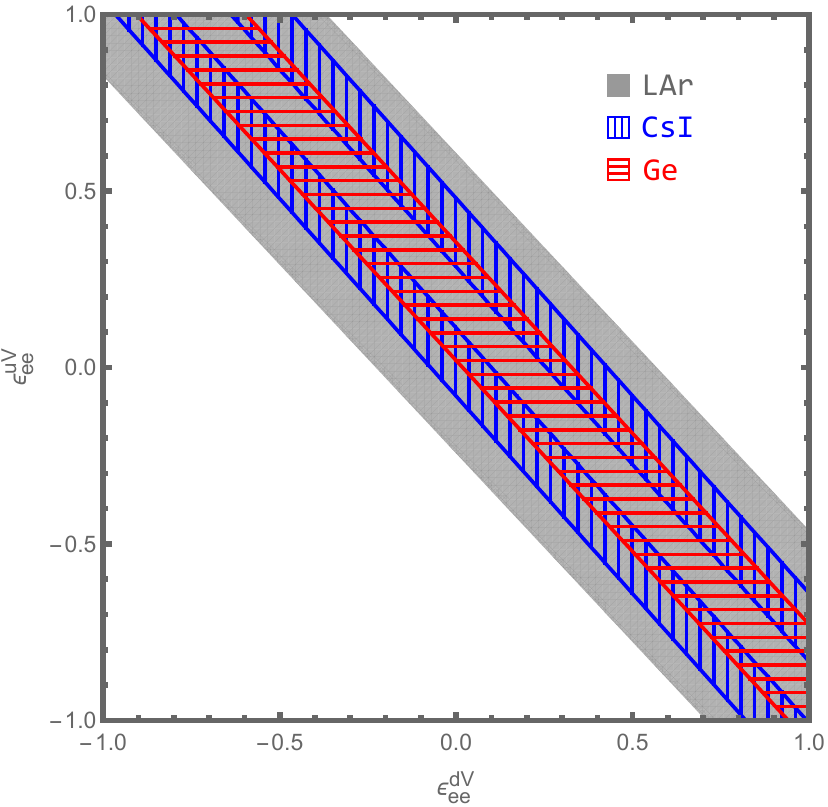}
\includegraphics[width=0.45\textwidth]{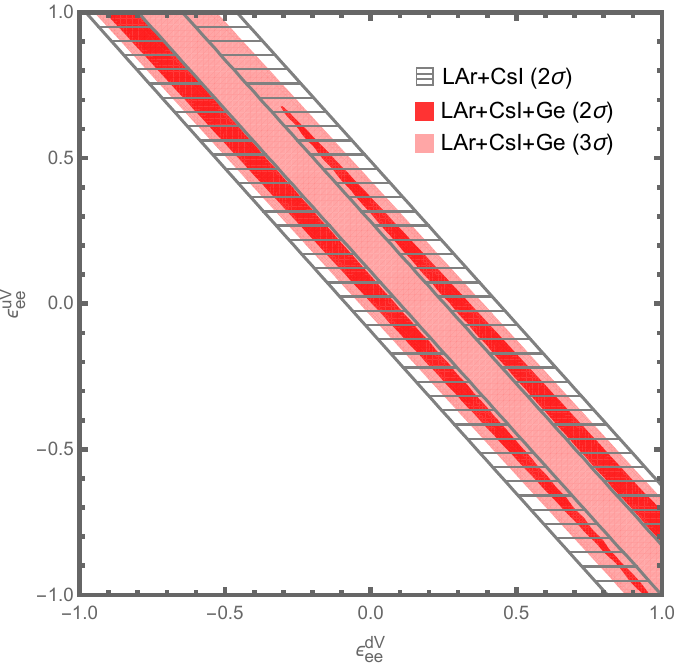}
\includegraphics[width=0.45\textwidth]{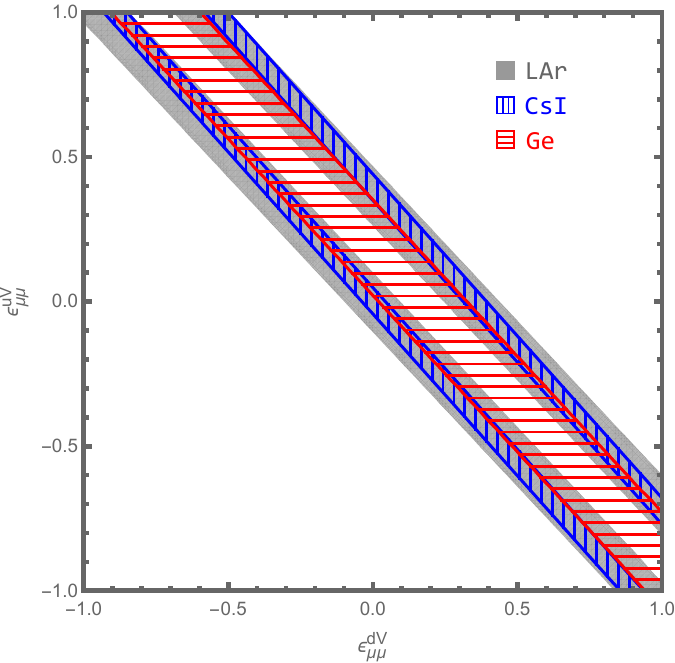}
\includegraphics[width=0.45\textwidth]{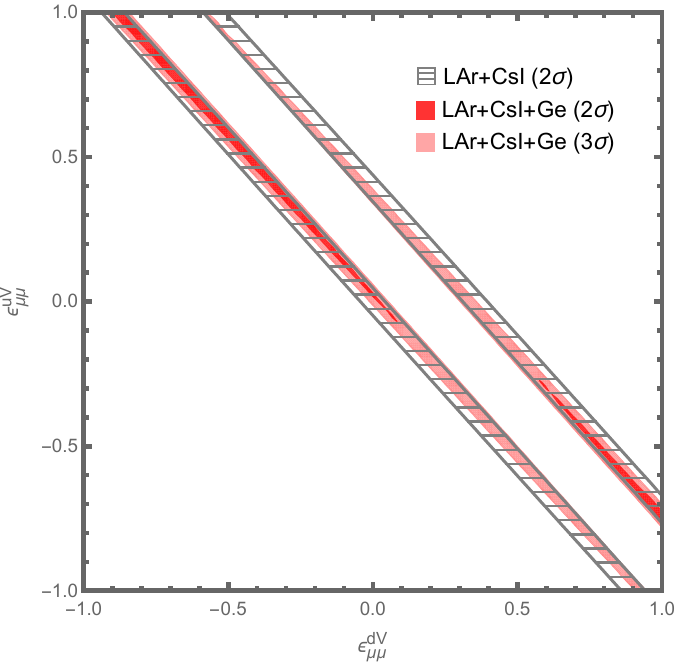}
\caption{Same as Fig.~\ref{fig:ee_uu_sinw}, except for ($\epsilon_{ee}^{dV}$, $\epsilon_{ee}^{uV}$) and ($\epsilon_{\mu\mu}^{dV}$, $\epsilon_{\mu\mu}^{uV}$).}
\label{fig:ee_uu}
\end{figure}

We investigate degeneracies between NSI parameters by fixing $\sin^2\theta_W=0.238$ and assuming that only the parameters in the selected parameter planes are nonzero.
In Fig.~\ref{fig:ee_uu} we show regions allowed in the ($\epsilon_{ee}^{dV}$, $\epsilon_{ee}^{uV}$) and ($\epsilon_{\mu\mu}^{dV}$, $\epsilon_{\mu\mu}^{uV}$) planes. 
 From Eq.~(\ref{eq:Q_NSI}) it is clear that the best fit parameters lie on a straight line given by 
\begin{align}[(g_V^p+\epsilon_{ee}^{dV}+2\epsilon_{ee}^{uV})Z+(g_V^n+2\epsilon_{ee}^{dV}+\epsilon_{ee}^{uV})N]=\pm[(g_V^p+\hat{\epsilon}_{ee}^{dV}+2\hat{\epsilon}_{ee}^{uV})Z+(g_V^n+2\hat{\epsilon}_{ee}^{dV}+\hat{\epsilon}_{ee}^{uV})N]\,,
\label{eq:slope}
\end{align}
where the hatted parameters are the best fits to the data. As Eq.~(\ref{eq:slope}) shows, the slope of the best fit line is related to the number of neutrons and protons in the nucleus. Since the best-fit lines for different elements have different slopes, the CsI, LAr and Ge data break degeneracies in the NSI parameters. From the right panels of Fig.~\ref{fig:ee_uu}, we see that the degeneracy in $\epsilon_{ee}^{dV}$ and $\epsilon_{ee}^{uV}$ ($\epsilon_{\mu\mu}^{dV}$ and $\epsilon_{\mu\mu}^{uV}$) is broken at the 2$\sigma$ CL. Due to the higher flux of $\numu$ compared to $\nue$, the constraints in the ($\epsilon_{\mu\mu}^{dV},\epsilon_{\mu\mu}^{uV}$) plane are stronger than in the ($\epsilon_{ee}^{dV},\epsilon_{ee}^{uV}$) plane, and the degeneracy between $\epsilon_{\mu\mu}^{dV}$ and $\epsilon_{\mu\mu}^{uV}$ is slightly broken even at 3$\sigma$. 

\begin{figure}
\centering
\includegraphics[width=0.45\textwidth]{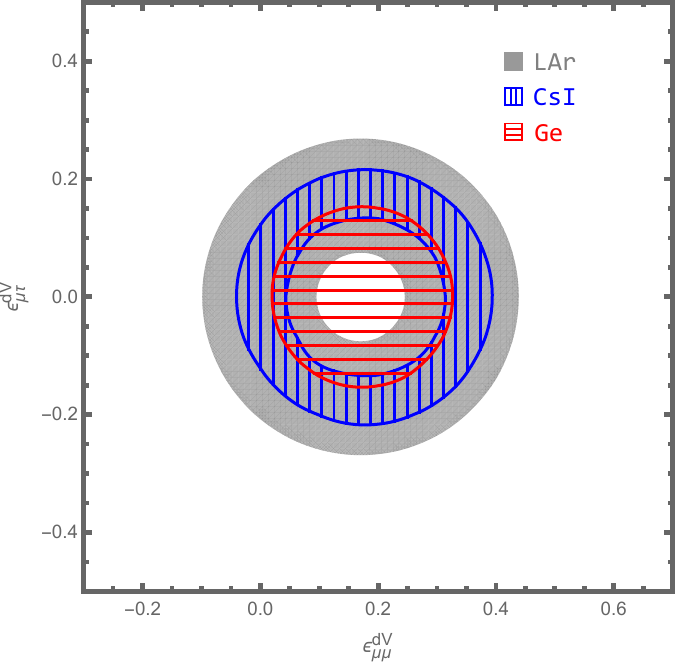}
\includegraphics[width=0.45\textwidth]{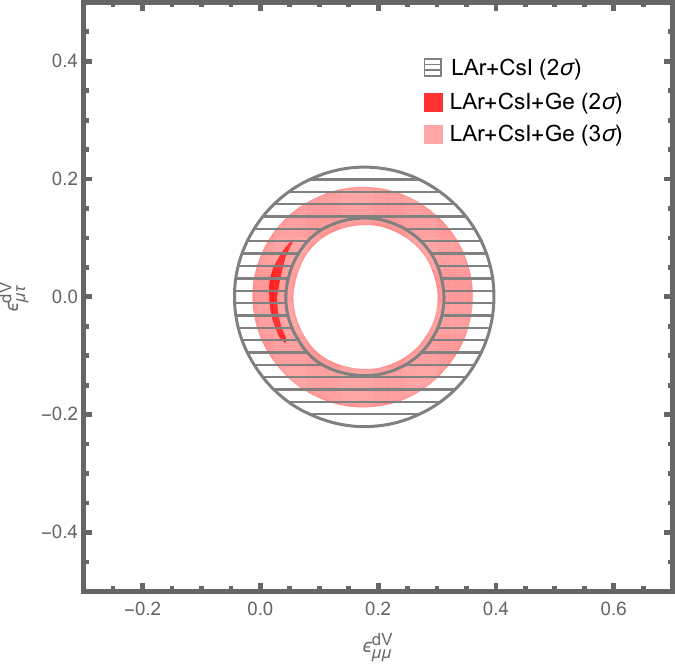}
\includegraphics[width=0.45\textwidth]{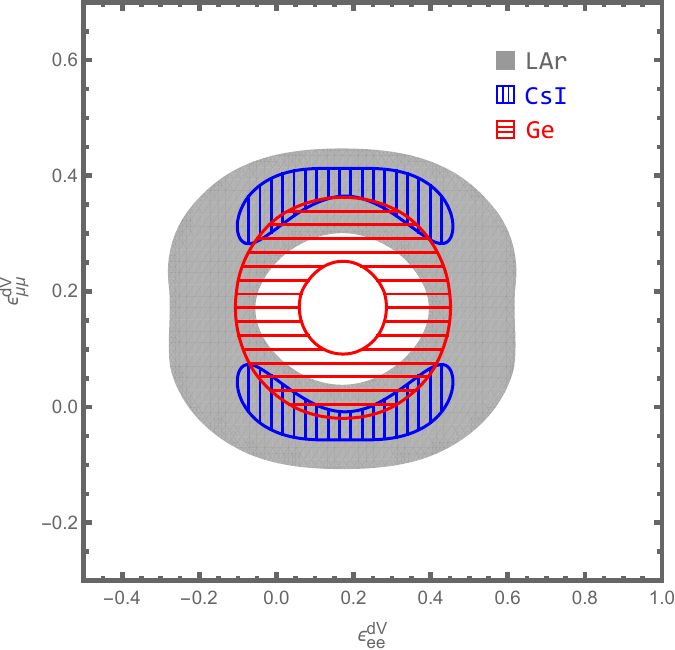}
\includegraphics[width=0.45\textwidth]{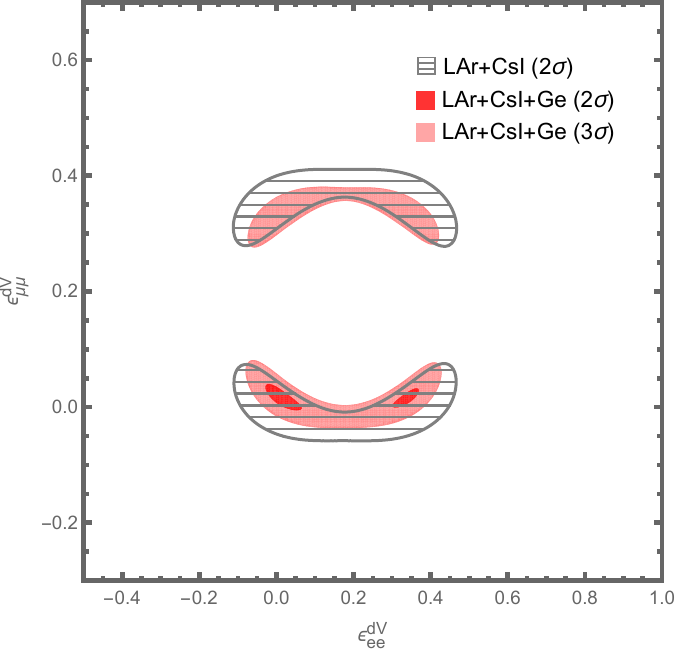}
\caption{Same as Fig.~\ref{fig:ee_uu_sinw}, except for ($\epsilon_{\mu\mu}^{dV}$, $\epsilon_{\mu\tau}^{dV}$) and ($\epsilon_{ee}^{dV}$, $\epsilon_{\mu\mu}^{dV}$).}
\label{fig:uu_ut uu_ee}
\end{figure}

We show allowed regions in the ($\epsilon_{\mu\mu}^{dV}, \epsilon_{\mu\tau}^{dV}$) and ($\epsilon_{ee}^{dV}, \epsilon_{\mu\mu}^{dV}$) planes in Fig.~\ref{fig:uu_ut uu_ee}.  
The shapes of the regions can be obtained from Eq.~(\ref{eq:Q_NSI}). For ($\epsilon_{\mu\mu}^{dV}$, $\epsilon_{\mu\tau}^{dV}$), the locus of best fit points is a circle  given by
\begin{align}
[(g_V^p+\epsilon_{\mu\mu}^{dV})Z+(g_V^n+2\epsilon_{\mu\mu}^{dV})N]^2+(\epsilon_{\mu\tau}^{uV}Z+2\epsilon_{\mu\tau}^{dV}N)^2=[(g_V^p+\hat{\epsilon}_{\mu\mu}^{dV})Z+(g_V^n+2\hat{\epsilon}_{\mu\mu}^{dV})N]^2+(\hat{\epsilon}_{\mu\tau}^{uV}Z+2\hat{\epsilon}_{\mu\tau}^{dV}N)^2\,.
\end{align}
Since the center and radius of the circle is different for different detector elements, the combined analysis breaks the degeneracy. From the upper right panel of Fig.~\ref{fig:uu_ut uu_ee}, we see that values of 
($\epsilon_{\mu\mu}^{dV}, \epsilon_{\mu\tau}^{dV}$) allowed at $2\sigma$ occupy a small region near the SM value $(0, 0)$, and those at 3$\sigma$ form a ring. 

For ($\epsilon_{ee}^{dV}$, $\epsilon_{\mu\mu}^{dV}$), we find that the minimum $\chi^2$ occurs at four points, obtained by solving  
\begin{align}
[(g_V^p+\epsilon_{ee}^{dV})Z+(g_V^n+2\epsilon_{ee}^{dV})N]&=\pm[(g_V^p+\hat{\epsilon}_{ee}^{dV})Z+(g_V^n+2\hat{\epsilon}_{ee}^{dV})N]\,,
\nonumber\\
[(g_V^p+\epsilon_{\mu\mu}^{dV})Z+(g_V^n+2\epsilon_{\mu\mu}^{dV})N]&=\pm[(g_V^p+\hat{\epsilon}_{\mu\mu}^{dV})Z+(g_V^n+2\hat{\epsilon}_{\mu\mu}^{dV})N]\,.
\end{align}
Since the locations of the four points depend on the detector elements, a joint analysis of the datasets breaks the degeneracy.
The bottom right panel of Fig.~\ref{fig:uu_ut uu_ee} shows that the 2$\sigma$ allowed parameter space in the ($\epsilon_{ee}^{dV}, \epsilon_{\mu\mu}^{dV}$) plane consists of two small disconnected regions, with the larger one near the SM value $(0, 0)$.

\section{Summary}

The COHERENT experiment has detected elastic neutrino-nucleus scattering in CsI, LAr and Ge. We analyzed a combination of these datasets to check if the scattering
is in fact coherent, and find that the cross section is proportional to $N^2$ to within $1.5\sigma$.
We also used the combined data to place constraints on the the weak mixing angle in conjunction with NSI parameters.  We find that current COHERENT data can effectively break degeneracies between various parameters at the 2$\sigma$ level. The Ge data are instrumental in significantly reducing the allowed NSI parameter space. 


\vspace{0.1in}
{\it Acknowledgments.} 
J.L. and J.Z. are supported by the National Natural Science Foundation of China under Grant No. 12275368 and the Fundamental Research Funds for the Central Universities, Sun Yat-Sen University under Grant No. 24qnpy116. D.M. is supported by the U.S. DoE under Grant No. de-sc0010504.


\bibliography{ref}

\end{document}